\documentclass[twocolumn,showpacs,preprintnumbers,amsmath,amssymb]{revtex4}
\usepackage{tabularx,graphicx}\begin{document}
\newcommand{\beq}{\begin{equation}}
\newcommand{\eeq}{\end{equation}}
\newcommand{\beqn}{\begin{eqnarray}}
\newcommand{\eeqn}{\end{eqnarray}}
\newcommand{\bmath}{\begin{subequations}}
\newcommand{\emath}{\end{subequations}}
\newcommand{\rb}{\bar{r}}
\newcommand{\bk}{\bold{k}}
\newcommand{\bkp}{\bold{k'}}
\newcommand{\bq}{\bold{q}}
\newcommand{\bkb}{\bold{\bar{k}}}
\newcommand{\br}{\bold{r}}
\newcommand{\brp}{\bold{r'}}
\newcommand{\vp}{\varphi}

\title{Kinetic energy driven superconductivity, the origin of the Meissner effect, and the reductionist frontier}
\author{J. E. Hirsch }
\address{Department of Physics, University of California, San Diego\\
La Jolla, CA 92093-0319}

\begin{abstract} 
Is superconductivity associated with a lowering or an increase of the kinetic energy of the charge carriers?
Conventional BCS theory predicts that the kinetic energy of  carriers increases in the transition from the normal to the
superconducting state. However,  substantial  experimental evidence obtained  in recent years indicates that in at least  some superconductors the opposite occurs.
Motivated in part by these experiments many novel mechanisms of superconductivity have recently been proposed where the transition to superconductivity is associated with a lowering of the kinetic energy of
the carriers. However none of these proposed unconventional mechanisms  explores the fundamental 
reason for kinetic energy lowering nor its wider implications. 
Here I  propose that kinetic energy lowering is  at the root of the Meissner effect, the most fundamental property of superconductors. The physics can be 
understood at the level of a single electron atom: kinetic energy lowering and enhanced diamagnetic susceptibility are intimately connected.
We propose that this connection extends to superconductors because they are, in a very real sense, `giant atoms'. 
According to the theory of hole superconductivity, 
superconductors expel negative charge from their interior driven by kinetic energy lowering and in the process expel any magnetic field lines
present in their interior.  Associated with this we predict  the existence of a macroscopic electric field in
the interior of superconductors and the existence of macroscopic quantum zero-point motion in the form of  a spin current in the ground state of
superconductors (spin Meissner effect).  
In turn, the understanding of the role of kinetic energy lowering in superconductivity suggests a new way to understand 
 the fundamental origin of kinetic
energy lowering in quantum mechanics quite generally. This provides a new understanding of `quantum pressure', the stability of matter and the origin of 
fermion anticommutation relations, it  leads to the prediction  that spin currents exist in the ground state of
aromatic ring molecules, and that the electron wave function is double-valued, requiring a reformulation of conventional quantum mechanics.

   \end{abstract}
\maketitle 
\section{introduction}
The first researcher to suggest that kinetic energy lowering may be  at the root of the phenomenon of superconductivity was 
(to this author's  knowlege) F. London. In the preface
to his 1950 book ``Superfluids'' he
writes: ``According to quantum
theory the most stable state of any system is $not$ a state of $static$ $equilibrium$
in the configuration of lowest potential energy. It is rather a
kind of $kinetic$ $equilibrium$ for the so-called zero point motion, which
may roughly be characterized as defined by the minimum average total (potential + kinetic) energy...''. Later on he writes
``It is not necessarily a configuration
close to the minimum of the potential energy (lattice order) which is the
most advantageous one for the energy balance, since by virtue of the
uncertainty relation the kinetic energy also comes into play. If the
resultant forces are $sufficiently$ $weak$ and act between $sufficiently$ $light$
$particles$, then the structure possessing the smallest total energy would
be characterized by a good economy of the kinetic energy...'', ... ``it would be the
outcome of a quantum mechanism of macroscopic scale.''

This remarkable insight was not incorporated in the development of the conventional BCS theory of superconductivity: BCS theory 
predicts that the kinetic energy of carriers $increases$ in the transition from the normal to the superconducting state. This is because the occupation
of momentum space single-particle states is spread out by the energy gap and as a consequence states of higher kinetic energy that were unoccupied in the
normal state at zero temperature become partially occupied in the superconducting ground state. Thus, the 
``good economy of the kinetic energy'' expected by London in the superconducting state was not realized in the first successful microscopic theory of superconductivity.

Experimental evidence in recent years however suggests that at least in some materials superconductivity $is$  associated with lowering of kinetic energy of the carriers. 
Early evidence for a change in high frequency optical absorption in cuprates upon the transition to superconductivity was reported by 
Dewing and Salje\cite{dewing} and Fugol et al\cite{fugol}. In 1999, Basov and coworkers\cite{basov} reported a violation of the 
Ferrell-Glover-Tinkham optical sum rule for superconductors in c-axis conduction in the cuprates, 
manifested in enhanced optical spectral weight at low frequencies in the superconducting state,
and shortly thereafter Van der Marel and coworkers\cite{marel}
and Santander and coworkers\cite{santander}
reported results indicating such a violation for in-plane conduction.
Subsequent experiments confirmed these observations for the underdoped regime of 
various high $T_c$ cuprates\cite{carboneexp,carbone2exp,deutscherexp}, and recently similar observations have
been reported for an iron-arsenide compound\cite{keimer}. Such optical effects are expected if the kinetic energy of the charge carriers is lowered in the superconducting state\cite{apparent,frank}. A correlation between suppression of 
low frequency optical spectral weight) (which is associated with $high$ kinetic energy) 
in the normal state and superconductivity has been noted in a wide variety of systems\cite{drude1, drude2}

Following these experimental developments (and in a few cases even before them)  it was pointed out that in several  models proposed to describe superconductivity induced by novel electronic mechanisms,
the kinetic energy of carriers is lowered in the transition to 
superconductivity\cite{k1,k1pp,k1p,k2,k3,k4,k5,k6,k7,k8,k9,k10,k11,k12,k13,k14,k15} , and it was argued that the experimental findings 
mentioned above lend support to these
models to describe unconventional superconductivity in various materials.

In this paper we want to analyze the physics of kinetic energy lowering from a fundamental point of view, starting from the physics of a single atom.
I will argue that the fundamental physics of kinetic energy lowering found in a single atom 
manifests itself in $only$ $one$ of the many theories proposed to describe superconductivity driven by kinetic energy lowering, namely
the   theory of hole superconductivity\cite{holetheory} and its associated models,  generalized Hubbard model with correlated hopping\cite{corrhop}, electron-hole asymmetric polaron models\cite{pol} and
dynamic Hubbard models\cite{dynhub}. The theory of hole superconductivity is the only theory that $predicted$ superconductivity through kinetic energy  lowering, and how 
this physics would
show up in optical properties,  many years 
before its experimental discovery\cite{apparent}.

Why should the physics of a single atom be relevant to the understanding of the supposedly complicated many body phenomenon that is superconductivity? Examples abound in physics where complicated
systems exhibit in essence the same properties as simpler systems (otherwise we would have little hope of making progress). For the topic of
interest here, superconductors display quantum coherence
at a macroscopic scale. It is natural to expect that they will share essential properties with the simplest systems we know that exhibit quantum coherence, i.e. atoms. The view that
superconductors are ``giant atoms'' was very prevalent in the past\cite{giant1,giant2,giant3,giant4}  after the London discovery that the diamagnetic response of superconductors is their more fundamental property (as opposed to
zero resistivity) and mimics the diamagnetism of atoms. I will argue here that superconductors share  more properties with atoms than originally suspected\cite{giantatom}. 
In particular, that they exhibit charge inhomogeneity as well as quantum zero-point motion at the macroscopic level, just as atoms do at the microscopic level.

Superconductors are `giant atoms'  and hence exhibit many properties of the microscopic world, but at the same time they exist in the macroscopic world.
Thus, in the spirit of Bohr's correspondence principle, their physics should be understandable $both$ from a microscopic quantum and from a 
macroscopic classical point of view. I will argue that identification of the $forces$ (a macroscopic concept) at play in the transition to superconductivity
is of great help in  understanding  the true nature of superconductivity. The conventional BCS theory does not address this issue and for that reason I argue cannot
explain the most fundamental phenomenon associated with superconductivity, the Meissner effect\cite{japan1}.

Finally, if superconductors are giant atoms, learning about the physics of superconductors may teach us something about the physics of the microscopic world  that we
didn't know before. In particular, I propose  that an understanding of the role of kinetic energy lowering in superconductivity can teach us why kinetic energy lowering
exists in the microscopic realm. We will find that kinetic energy lowering is essentially tied to angular momentum,
which is different from the conventional understanding arising from quantum mechanics. Thus I argue that  the study of superconductivity gives us insights that may change our understanding at the `reductionist frontier'\cite{reductionist}.

\section{kinetic energy driven `superconductivity' in a single   atom}

\subsection{One electron}

The single electron in a hydrogen-like ion has a phase-coherent wavefunction, just as the wavefunction of a macroscopic superconductor.
I argue here that  the single electron atom illustrates much of the essential physics of 
superconductivity, since it exhibits  the physics of what kinetic energy lowering means in its simplest form.

The Hamiltonian is
\beq
H=-\frac{\hbar^2}{2m_e}\nabla^2-\frac{Ze^2}{r} \equiv H_{kin} + H_{pot}
\eeq
with $(-Ze)$ the ionic charge. Consider the  wavefunction
\beq
\psi _{\rb} (r)=(\frac{1}{\bar{r}^3 \pi})^{1/2}e^{-r/\rb}
\eeq
The most probable radial position for an electron described by this wavefunction is $r=\rb$. 
The expectation values of kinetic and potential energies with this wavefunction are
\bmath
\beq
E_{kin}(\rb)=<H_{kin}>=\frac{\hbar^2}{2m_e \rb^2}
\eeq
\beq
E_{pot}(\rb)=<H_{pot}>=-\frac{Ze^2}{\rb}
\eeq
\emath
and the minimum total energy results for $\rb=r_0=a_0/Z$, with  $a_0=\hbar ^2 / (m_e e^2) $ the Bohr radius, which is of course the ground state energy of
Eq. (1).

 \begin{figure}
\resizebox{8.5cm}{!}{\includegraphics[width=7cm]{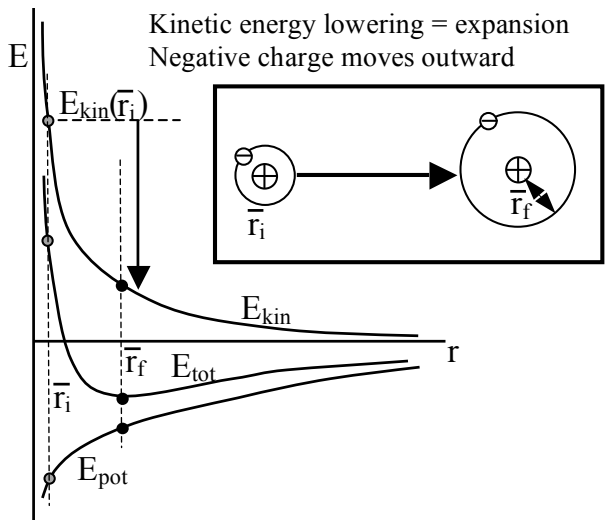}}
\caption { Kinetic energy lowering in the singe electron atom as the orbit expands. 
The energies for the small orbit of radius $\rb_i$ are shown as grey circles, those of the large radius $\rb_f$ as black circles }
\label{figure1}
\end{figure}

Suppose the nucleus of this ion has initially charge $Z_i$ and the electron is in its ground state, Eq. (2) with $\bar{r}_i=a_0/Z_i$.
Assume at time $t=0$ one or several protons in the nucleus undergo inverse beta decay and convert into neutrons, and 
 the nuclear charge becomes $Z_f<Z_i$. At $t=0^+$ the electron will 
still be described by the wavefunction Eq. (2) with $\bar{r}   = \rb _i  =a_0/Z_i$ (sudden approximation), and will evolve towards the ground state of the new Hamiltonian,
i.e. Eq. (2) with $\bar{r}_f=a_0/Z_f$ through spontaneous emission of photons. Since $\bar{r}_i<\bar{r}_f$  the electronic wavefunction will $expand$. We have  from Eq. (3)
\bmath
\beq
E_{kin}(\bar{r}_f)<E_{kin}(\bar{r}_i)
\eeq
\beq
E_{pot}(\bar{r}_f)>E_{pot}(\bar{r}_i)
\eeq
\beq
E_{total}(\bar{r}_f)<E_{total}(\bar{r}_i)
\eeq
\emath
So the initial state has high kinetic energy and low potential energy, and in the process of expanding the wavefunction to lower the total energy the kinetic energy is lowered and the potential
energy is raised  (by a lesser amount), and negative charge moves outward. This is shown schematically in Figure 1.
Charge separation occurs at a cost in Coulomb electrostatic energy, driven by 
an `electromotive force'\cite{emf}, namely kinetic energy lowering.

The orbital (Larmor diamagnetic) susceptibility is
\bmath
\beq
\chi_{Larmor}=-\frac{e^2}{6m_e c^2} <r ^2>=-\frac{e^2}{4m_e c^2}<r_\bot ^2> 
\eeq
for the wavefunction Eq. (2), where $<r_\bot ^2>=(2/3)<r^2> =2\bar{r}^2$ denotes the average of the square radial distance in the
plane perpendicular to the magnetic field. The orbital magnetic moment of the atom in the presence of an external magnetic field $\bold{B}$ is
\beq
\bold{m}=\chi_{Larmor}\bold{B}
\eeq
\emath
As the wavefunction expands from $\bar{r}_i$ to $\bar{r}_f>\bar{r_i}$ the magnitude of $\chi_{Larmor}$ increases. If an external magnetic field is present, an initially small orbital magnetic moment pointing antiparallel to the field  increases in magnitude:
\beq
\Delta\bold{m}=-\frac{e^2}{4m_e c^2}(<r_\bot ^2>_f-<r_\bot ^2>_i)\bold{B}
\eeq
and gives rise to an increasingly larger magnetic field in direction opposite to that of the applied field in the region inside the orbit.  In Eq. (6), 
$<r_\bot ^2>_f$ and $<r_\bot ^2>_i$
  denote the average of the square radial distance in the
plane perpendicular to the magnetic field in the final and initial states.

 \begin{figure}
\resizebox{8.5cm}{!}{\includegraphics[width=7cm]{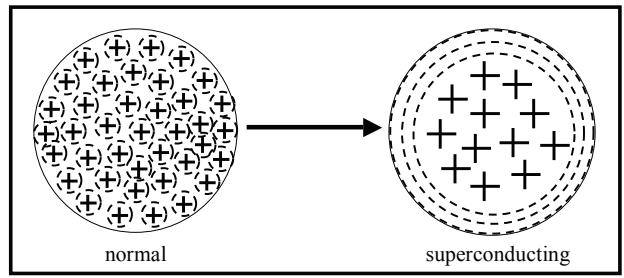}}
\caption {In a kinetic energy driven superconductor, the expansion of the orbits associated with kinetic energy lowering gives rise to negative
charge expulsion and macroscopic charge inhomogeneity, with more negative charge near the surface and more positive charge in the
interior. The potential energy is lower in the normal state where the charge is uniformly distributed.}
\label{figure1}
\end{figure} 

The physics just described is the physics of kinetic energy driven superconductivity as described by the theory
of hole superconductivity: kinetic energy is lowered\cite{apparent,kinetic},
potential energy increases, the wavefunction expands\cite{holeelec2} and negative charge moves radially outward\cite{chargeexp} giving rise to 
a macroscopically inhomogeneous charge distribution as shown schematically in Figure 2. Associated with this, the diamagnetic response increases, and 
if an external magnetic field is present as the transition to superconductivity occurs
 an orbital magnetic moment opposite to the external magnetic field grows in magnitude. To obtain the perfect
diamagnetism of superconductors, the diamagnetic susceptibility of $n$ electrons per unit volume has to take the value $-1/(4\pi)$, hence  
\beq
-\frac{1}{4\pi}=-\frac{ne^2}{4m_e c^2}<r_\bot^2>
\eeq
which will be the case when the radial dimension of the orbit is such that $\sqrt{<r_\bot^2>}=2\lambda_L$ with $\lambda_L$  the London penetration depth   given by\cite{tinkham}
\beq
\frac{1}{\lambda_L^2}=\frac{4\pi n e^2}{m_ec^2}    .
\eeq
The fact that the perfect diamagnetism of superconductors can be understood if  electrons reside in orbits of radius of order $\lambda_L$ was pointed out by
Frenkel\cite{frenkel}, Smith and Wilhelm\cite{smith} and  Slater\cite{slater} in 1933-1937,  but is not part of conventional BCS theory.
Note  that what $drives$ the expansion of the orbits is not the external magnetic field but kinetic energy lowering. Thus we would expect it to occur independently of 
whether an external magnetic field is or is not present in the transition to superconductivity.

Note also that in the `atoms' we are considering, whether small or large, the diamagnetic magnetic moment that  grows  as the wavefunction expands in an external
magnetic field  can be understood as arising from
the magnetic Lorentz force $\bold{F}=(e/c)\bold{v}\times\bold{B}$ acting on the radially outgoing electron. The classical equation of motion is
\beq
m_e\frac{d\bold{v}}{dt}=\frac{e}{c}\bold{v}\times{B}+\bold{F}_r
\eeq
where the second term on the right  is the radial force arising from kinetic energy lowering in the expanding orbit. From Eq. (9),
\beq
\bold{r}\times\frac{d\bold{v}}{dt} = \frac{e}{m_e c}\bold{r}\times(\bold{v}\times\bold{B})
\eeq
where $\bold{r}$ is in the plane perpendicular to $\bold{B}$. Hence $\bold{r}\times(\bold{v}\times\bold{B})=-(\bold{r}\cdot\bold{v})\bold{B}$, and
\beq
\frac{d}{dt}(\bold{r}\times \bold{v})= -\frac{e}{m_e c}( \bold{r}\cdot \bold{v})\bold{B}=-\frac{e}{2m_e c} (\frac{d}{dt}r^2)\bold{B}
\eeq
so that in expanding from $\rb _i$ to $\rb _f$ 
\beq
\rb _f v_f-\rb _i v_i=-\frac{e}{2m_e c} (\rb _f^2-\rb _i^2)B
\eeq
and the change in the magnetic moment  
\beq
\bold{m}=\frac{e}{2c}\bold{r}\times\bold{v}
\eeq
is
\beq
\Delta \bold{m}=-\frac{e^2}{4m_e c^2}(\rb_f^2-\rb_i^2)\bold{B}
\eeq
in agreement with Eq. (6).  Thus we arrive at  a   $dynamical$ understanding of the Meissner effect\cite{sm}, both from a quantum and a classical point of view, as
being intimately tied to the lowering of kinetic energy that occurs when the electronic orbit expands and negative charge moves outward.

Note also that the electron-ion interaction $E_{pot}$ (Eq. (3b)) that works $against$ orbit expansion and kinetic energy lowering is proportional to the ionic charge $Z$. To the 
extent that this physics is relevant to superconductivity, we would expect superconductivity to be favored in systems where the effective ionic charge is as small
as possible, which corresponds to the situation where the atoms are $negatively$ $charged$ $anions$\cite{holesc}.

In summary, we have shown in this section that a fundamental relationship exists between kinetic energy lowering, increased diamagnetism, orbit expansion and outward motion of
negative charge. It is only natural to expect that
since superconductors undergo a giant increase in their diamagnetism in the transition to superconductivity, kinetic energy lowering, orbit expansion and negative charge expulsion should also take place. None of this   however is described by  conventional BCS theory.

Furthermore, none of the numerous other `kinetic energy driven' superconductivity mechanisms discussed in the literature\cite{k1,k1pp,k1p,k2,k3,k4,k5,k6,k7,k8,k9,k10,k11,k12,k13,k14,k15}
 contain any of  this physics.  For example, in the review article ``Concepts in High Temperature Superconductivity''\cite{k1p} the authors
state that in high temperature superconductors ``the condensation is driven by a lowering
of kinetic energy'',  that   ``The Spin Gap Proximity Effect Mechanism" provides
 ``a novel route to superconductivity through kinetic-energy driven pairing'', and that  ``{\it ItÕs all about kinetic energy}''. Yet their mechanism contains none of
 the fundamental physics of kinetic energy lowering  exhibited by the single-electron atom, and neither it nor any of the other `kinetic energy' mechanisms of superconductivity proposed\cite{k1,k1pp,k1p,k2,k3,k4,k5,k6,k7,k8,k9,k10,k11,k12,k13,k14,k15}  has
 anything to say about the relation between kinetic energy lowering and the  origin of the Meissner effect.

\subsection{Two electrons}

The two-electron atom provides us with  additional insight into the physics of kinetic energy driven superconductivity. The Hamiltonian is
\beq
H=-\frac{\hbar^2}{2m_e}(\nabla_1^2+\nabla_2^2) - Ze^2(\frac{1}{r_1}+\frac{1}{r_2})-\frac{e^2}{|\bold{r}_1-\bold{r}_2|}
\eeq
We consider the simple  variational wavefunction
\beq
\Psi_{\rb} (\bold{r}_1,\bold{r}_2)=\psi_{\rb}  (r_1)\psi _{\rb} (r_2)  
\eeq
The different contributions to the two-electron atom energy with the wavefunction Eq. (16) are
\bmath
\beq
E_{kin}(\rb) =2\frac{\hbar^2}{2m_e \rb^2}
\eeq
\beq
E_{e-i}(\rb)= -2\frac{Ze^2}{\rb}
\eeq
\beq
E_{e-e}(\rb)=\frac{5}{8} \frac{e^2}{\rb}
\eeq
\emath
hence  orbital expansion (increasing $\rb$)  reduces both the kinetic energy and the electron-electron repulsion energy, and increases the electron-ion energy.
The electrostatic potential energy is the sum of the electron-ion ($E_{e-i}$) and electron-electron ($E_{e-e}$) Coulomb energies:
\beq
E_{pot}(\rb)=2e^2[\frac{5}{16}-Z]\frac{1}{\rb}
\eeq
Clearly it requires $Z>5/16$ for the system to be stable (this is a necessary but not sufficient condition), hence in that regime, orbital expansion increases
the electrostatic energy (as it should because it is associated with  charge separation) but less so than it would in the absence of electron-electron interaction. The total energy
\beq
E_{total}(\rb)=2e^2[\frac{5}{16}-Z]\frac{1}{\rb}+2\frac{\hbar^2}{2m_e \rb^2}
\eeq
is minimized by  $\rb=a_0/(Z-5/16)$. That is, the orbital expands from the radius it would have in the absence of electron-electron interaction, $\rb=a_0/Z$,
driven by the electron-electron Coulomb repulsion.

Thus, assuming that the electron wavefunction expands in the transition from  the normal to the superconducting state, one could say that the
transition is `driven' by $both$ kinetic energy lowering $and$ electron-electron Coulomb repulsion lowering, and opposed by the
electron-ion Coulomb attraction. But if we just consider kinetic energy
versus  total potential (electrostatic) energy we would say that the transition is `driven' by kinetic energy gain at a cost in potential energy.

Furthermore, the single particle energy per electron 
\beq
\epsilon_{s.p.}(\rb)=\frac{\hbar^2}{2m_e \rb^2} - \frac{Ze^2}{\rb}
\eeq
is minimum for $\rb=a_0/Z$, hence is higher for an `expanded' orbital. In other words, the electron in the expanded orbital occupies higher energy single-particle
states. Thus we conclude that if superconductivity is associated with expansion of the orbital
it will also involve electronic occupation of higher energy single-particle states than the normal state.

\section{kinetic energy in electronic energy bands}

The kinetic energy of electrons at the Fermi energy increases as the electronic band occupation increases. It is reasonable to expect that the most favorable regime for kinetic energy driven
superconductivity will be when electrons at the Fermi energy have highest kinetic energy in the normal state, which is the case when the band is almost full. This is shown schematically in Fig. 3.

 \begin{figure}
\resizebox{8.5cm}{!}{\includegraphics[width=7cm]{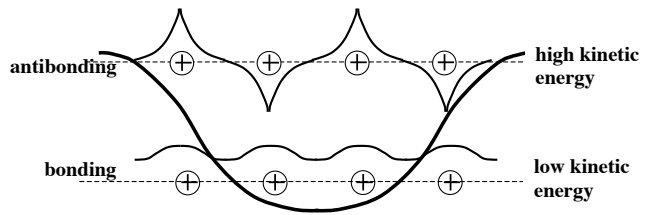}}
\caption {Electronic energy states in a band (schematic). The states near the bottom of the band (electron states) are smooth (low kinetic energy) and have high charge
density in the region between the ions (bonding). The states near the top of the band (hole states) are wiggly (high kinetic energy) and have low charge density
in the region between the ions (antibonding).}
\label{figure1}
\end{figure} 

Beyond single-electron physics, an intimate relation exists between kinetic energy and electron-hole asymmetry when the electron-electron
interaction is considered\cite{corrhop}. In a tight binding model with all nearest neighbor matrix elements of the Coulomb interaction included (generalized Hubbard model), the $only$ Coulomb interaction
term in the Hamiltonian that breaks electron-hole symmetry is the correlated hopping term\cite{bondch}, of the form:
\beq
H_{\Delta t}=\Delta t \sum_{<ij>\sigma}(n_{i,-\sigma}+n_{j,-\sigma})  (c_{i \sigma}^\dagger c_{j\sigma}+h.c.)
\eeq
This has a form closely related to  the single particle kinetic energy\cite{coulombattr}
\beq
H_t=-t _0\sum_{<ij>\sigma}  (c_{i \sigma}^\dagger c_{j\sigma}+h.c.)
\eeq
and it gives rise to an effective hopping amplitude $t=t_0-n\Delta t$, with $n$ the band occupation. The interaction Eq. (21) is repulsive near the bottom of the band where
the wavefunction is smooth and the expectation value of $c_{i \sigma}^\dagger c_{j\sigma}$ is positive, and attractive where the wavefunction changes sign in going from a site to
a neighboring site, i.e. near the top of the band.

\section{kinetic energy lowering in hole superconductivity}

\begin{figure}
\resizebox{8.5cm}{!}{\includegraphics[width=7cm]{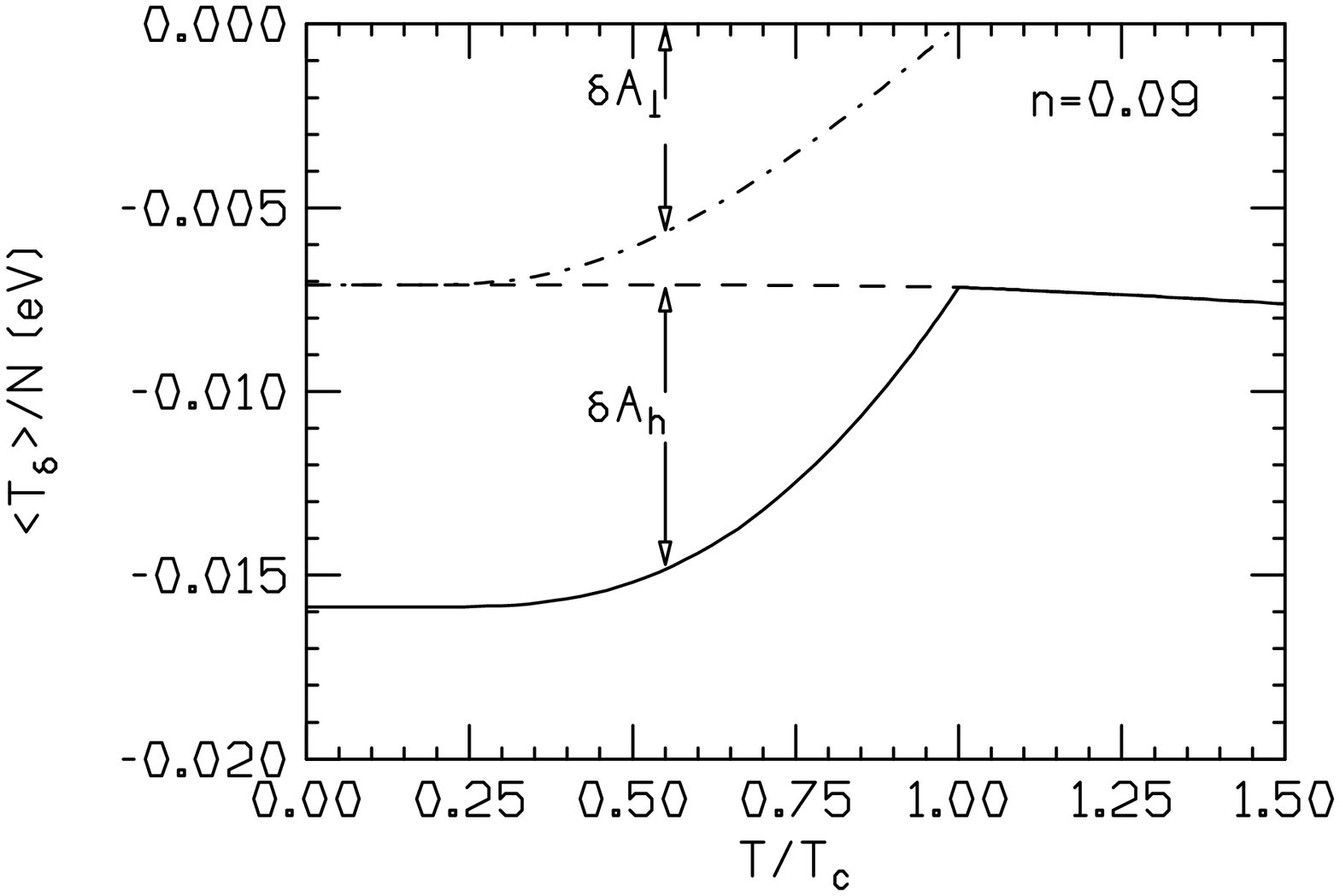}}
  \caption{Average kinetic energy per lattice site vs. temperature for Hamiltonian parameters $U=5$, $V=0.65$, $\Delta t=0.47$ (from Ref. 
  \cite{apparent}, Fig. 2a). The full line shows
  the total kinetic energy per particle. The terms $\delta A_l$ and $\delta A_h$ refer to the `missing area' in optical absorption at low and high frequencies respectively
  (see Ref. \cite{apparent} for a full discussion).}
\end{figure}

In a BCS treatment of the tight binding Hamiltonian with correlated hopping as well as on-site ($U$) and nearest neighbor repulsion ($V$)  it is found that superconductivity occurs 
even when the net interaction strength is repulsive, because of the energy dependence of the gap function\cite{corrhop} that results from the 
interaction term Eq. (21). In addition, it is found that the kinetic energy is
lowered as the system enters the superconducting state\cite{apparent,ninetynine}. This is shown in Fig. 4 for one case. The lowering of kinetic energy is closely related to the anomalous
behavior in optical absorption found in refs. \cite{fugol,marel,santander} and predicted in ref.  \cite{apparent}. Qualitatively, the lowering of kinetic energy upon pairing is
easily understood from the fact that the hopping amplitude of a hole carrier $increases$ by $\Delta t$ when another carrier is present in one of the two sites
involved in the hopping process according to Eq. (21).

These results obtained within BCS theory are expected to be accurate because BCS theory becomes exact in the limit of small carrier concentration\cite{eagles}, which is the
regime of interest here. This has been verified by exact diagonalization studies of this\cite{parola,lin} and related\cite{dynh} Hamiltonians.

\begin{figure}
\includegraphics[height=.10\textheight]{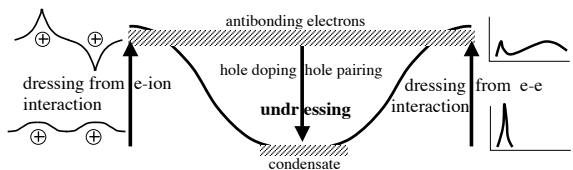}
  \caption{As the Fermi level rises in a band,   electrons at the Fermi energy 
 become  dressed due to electron-electron interactions which modify the free particle
  spectral function (as shown schematically on the right of the figure), and due to electron-ion interactions which modify the
  free-electron wave function (as shown on the left side of the figure). Pairing effectively lowers the position
  of the Fermi level and causes the carriers at the Fermi energy to 'undress' and become
  free-electron-like.}
\end{figure}

But this is not the whole story. The profound connection between kinetic energy lowering and hole superconductivity emerges from several other angles.
Carriers in the normal state are found to be highly `dressed' due to both the effect of electron- electron\cite{holeelec} $and$ the electron-ion\cite{holeelec2} interaction when the band is 
almost full. When carriers pair and the system becomes superconducting, carriers `undress'\cite{undressing}. The hole character of a carrier in the normal state of the almost full band arises from the 
fact that it is `dressed' by the electron-ion interaction and has  a short wavelength that is sensitive to the electron-ion potential. When carriers pairs and 'undress',
they no longer `see' the discrete electron-ion potential because their wavelength becomes much larger than the lattice spacing\cite{holeelec2}. Holes turn into electrons,
and the wave function of carriers goes from being highly `wiggly' indicating high kinetic energy   to smooth, indicating low kinetic energy, as shown schematically in Fig. 5.
The right-hand side of Fig. 5 shows schematically the spectral function which is highly incoherent in the normal state when the band is almost full and carriers are
highly dressed and becomes
coherent when carriers `undress'.

\section{superconductor as a giant atom and charge redistribution}

The above considerations lead to the conclusion that electrons in superconductors do not `see' the lattice periodicity but rather move in a uniform positive background: the 
superconductor is a giant `Thomson atom'\cite{giantatom}. It is natural to expect that the charge density in such a system will be macroscopically inhomogeneous, just as it is microscopically inhomogeneous
in an ordinary atom. Electrons  expand their wavefunction to lower their kinetic energy and this is associated with expulsion of negative charge from the interior to the surface, just as for the
ordinary atom discussed in section II. The result is an excess of negative charge near the surface of the superconductor, as shown schematically in Figure 2.

The electrodynamics of superconductors in this framework is described\cite{electrodyn} by a modification of conventional London electrodynamics, of a form close to equations originally considered by the
London brothers\cite{giant1}  but later abandoned by them. The electric potential and charge density are related by the equation
\beq
\rho- \rho _0=- \frac{1}{4\pi\lambda _L^2} (\phi-\phi _0)
\eeq
which is  the fourth component of a four-dimensional relativistically covariant description:
\beq
J-J_0=-\frac{c}{4\pi\lambda_L^2}(A-A_0)
\eeq
where the four-vectors are given by 
\bmath
\beq
A=(\bold{A},i\phi)
\eeq
\beq
J=(\bold{J},ic\rho)
\eeq
\emath
and
\bmath
\beq
A_0=(0,i\phi_0)
\eeq
\beq
J_0=(0,ic\rho_0)
\eeq
\emath
The spatial part of Eq. (24) is the ordinary London equation relating current density and magnetic vector potential. Contrary to the conventional London equation where the
vector potential obeys the London / Coulomb gauge $\bold{\nabla}\cdot\bold{A}=0$, here the vector potential obeys the Lorentz gauge
\beq
\bold{\nabla}\cdot\bold{A}+\frac{1}{c}\frac{\partial \phi}{\partial t}=0
\eeq
consistent with the continuity equation
\beq
\bold{\nabla}\cdot\bold{J}+\frac{1}{c}\frac{\partial \rho}{\partial t}=0 .
\eeq
The uniform charge density $\rho_0$ in the interior gives rise to the potential $\phi_0$ through the usual electrostatic relation. 
Within a London penetration depth of the surface the charge density becomes negative and is denoted by $\rho_ -$. 

These electrodynamic equations predict that the screening length for electrostatic fields in superconductors is $\lambda_L$, much larger than the 
Thomas Fermi screening length of normal metals. This `rigidity' of the superconductor with respect to electric perturbations is qualitatively different from the prediction
of BCS theory that electric fields in superconductors are screened just like in normal metals.

The relation between $\rho_0$ and $\rho _-$ depends on
the geometry of the sample. In particular, for spherical, cylindrical and planar samples one has respectively
\bmath
\beq
\rho_0=-\frac{3\lambda_L }{R}\rho_-
\eeq
\beq
\rho_0=-\frac{2\lambda_L }{R}\rho_-
\eeq
\beq
\rho_0=-\frac{2\lambda_L }{D}\rho_-
\eeq
\emath
where $R$ is the radius in the spherical and cylindrical case, and D is the thickness in the planar case. In all cases, the electric field in the interior attains
its maximum value $E_m$ within a London penetration depth of the surface, given by
\beq
E_m=-4\pi \lambda_L \rho_-   .
\eeq 
From energetic considerations one deduces that $\rho_-$ and $E_m$ are independent of sample dimensions (for sample dimensions much larger than $\lambda_L$) while
$\rho_0$ decreases with sample size. 
 For more general geometries, the negative charge density $\rho_-$ near the surface is not uniform but a function of position.

\section{$2\lambda_L$ orbits and Meissner effect}
 
 Superconductivity can be understood semiclassically if the superfluid carriers reside in real space orbits of radius $2\lambda_L$. This can be seen in various ways. The angular
 momentum of superconducting carriers of density $n_s$ moving with speed $v_s$ within $\lambda_L$ of the surface of a cylinder of radius $R$ and unit height   is
 \beq
 L_s=[n_s(2\pi R) \lambda_L][m_e v_s R]
 \eeq
 where the first factor is the number of carriers within $\lambda_L$ of the surface and the second the angular momentum of each carrier (for $R>>\lambda_L$). $L_s$ can also be written as
 \beq
 L_s=[n_s(\pi R^2)][m_e v_s (2\lambda_L)]
 \eeq
 where the first factor is $all$ the superfluid carriers, and the second factor is the angular momentum of an electron in a circular orbit of radius $2\lambda_L$. The velocities 
 in the interior cancel out, and the net superfluid flow occurs only within $\lambda_L$ of the surface.
 
 In addition, the Larmor diamagnetic susceptibility of carriers of density $n_s$ in circular orbits of radius $r$ perpendicular to an applied magnetic field is
 \beq
 \chi_{Larmor}(r)=-\frac{n_s e^2}{4m_e c^2}r^2
 \eeq
 and with the London penetration depth given by Eq. (8),
 \beq
  \chi_{Larmor}(r=2\lambda_L)=-\frac{1}{4\pi}
  \eeq
  corresponding to perfect diamagnetism. The suggestion that the perfect diamagnetism of superconductors requires the carriers to be in orbits of radius of order
  $\lambda_L$ was made by several workers before BCS theory\cite{frenkel,smith,slater} but is not part of BCS theory.
  
  In the normal state, the carriers can be assumed to be in microscopic orbits of radius $k_F^{-1}$, with $k_F$ the Fermi wavevector, of order $\AA^{-1}$. Indeed, the
  Larmor diamagnetic susceptibility of carriers in such orbits is 
    \beq
 \chi_{Larmor}(r=k_F^{-1})=-\frac{n_s e^2}{4m_e c^2}k_F^{-2}=-\frac{1}{3}\mu_B^2g(\epsilon_F)
 \eeq
 with 
 \beq
 g(\epsilon_F)=\frac{3n_s}{2\epsilon_F}
 \eeq
  the density of states at the Fermi energy $\epsilon_F=\hbar^2k_F^2/2m_e$ and $\mu_B=e\hbar/2m_ec$ the Bohr magneton.
  Eq. (35) is the well-known expression for the small Landau orbital diamagnetism in the normal state.
  
  As the system goes superconducting, the electronic orbits will expand from radius $k_F^{-1}$ to radius $2\lambda_L$ driven by kinetic energy lowering,
  and the Lorentz force on the radially outgoing electron will impart it with the azimuthal velocity required to yield perfect diamagnetism when the orbit reaches
  radius $2\lambda_L$, as shown by Eqs. (2)-(14).    The $2\lambda_L$ orbits are highly overlapping, in contrast to the $k_F^{-1}$ orbits that are
  non-overlapping.

  \section{$2\lambda_L$ orbits and Spin Meissner effect}
  
  Consider the process of orbit expansion in the absence of a magnetic field. The  interaction of the moving electron magnetic moment 
  with the ionic positive background of charge density
  $|e|n_s$ (determined by charge neutrality) will impart the electron with an azimuthal velocity that depends on the spin orientation. 
  From the equation of motion one finds\cite{sm} that the azimuthal velocity of an electron that expanded to radius $\vec{r}$ is
    \beq
  \vec{v}_\sigma=-\frac{\pi e n_s \mu_B}{m_e c} \vec{\sigma}\times  \vec{r}
  \eeq
  and for radius $r=2\lambda_L$ Eq. (37) yields, using Eq. (8)
  \beq
   \vec{v}_\sigma^0=-\frac{\hbar}{4m_e \lambda_L} \vec{\sigma}\times \hat{r}
  \eeq
Just like for the Meissner effect, the `internal' velocities cancel out and the only macroscopic superfluid motion occurs within a London penetration depth of the surface.
In a cylinder, electrons of spin up and spin down parallel to the axis of the cylinder circulate clockwise and 
counterclockwise respectively as seen from the top of the cylinder, as shown in Fig. 6.
Thus we predict that a macroscopic spin current circulates near the surface of superconductors in the absence of applied fields.

It should be pointed out that many workers before the advent of BCS theory proposed that spontaneous currents exist in superconductors in the absence of 
applied fields. These include Bloch\cite{bloch}, Landau\cite{landau}, Frenkel\cite{frenkel}, Smith\cite{smith}, Born and Cheng\cite{born},
Heisenberg\cite{heisenberg} and Koppe\cite{koppe}. Notably, some of these proposals were made even before the discovery of the Meissner effect\cite{bloch,landau,frenkel}.
However these workers envisioned charge currents with 
different orientations in different domains to give rise to zero net macroscopic charge current,  rather than the spin current discussed here that does not require domains.

\begin{figure}
\resizebox{6.5cm}{!}{\includegraphics[width=7cm]{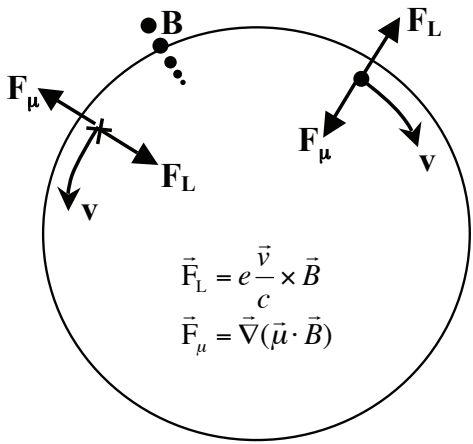}}
  \caption{Forces acting on electrons in the spin current in the presence of a magnetic field pointing out of the paper. For the electron with spin pointing
  into the paper, the Lorentz force Eq. (39) points inward and the gradient force Eq. (40) points outward. For the electron with spin pointing out of
  the paper the situation is reversed.}
\end{figure}

Strong corroborating evidence for the physics discussed here is the following argument: electrons circulating near the surface of a superconductor  with velocity $\vec{v}$ in the presence of a 
small magnetic field  will experience two radial forces: a Lorentz force due to the motion of the spin current,
\beq
\vec{F}_L=\frac{e}{c}\vec{v}\times\vec{B}
\eeq
and a gradient force pushing the electron magnetic moment that is parallel (antiparallel) to the magnetic field in the direction of higher (lower) magnetic field:
\beq
\vec{F}_\mu=\vec{\nabla}(\vec{\mu}\cdot\vec{B})
\eeq
as shown in Fig. 6. The condition that these forces exactly compensate each other
\beq
\frac{e}{c}\vec{v}\times\vec{B}=\vec{\nabla}(\vec{\mu}\cdot\vec{B})
\eeq
yields for the velocity the value Eq. (38), using the fact that the radial gradient of the magnetic field near the surface is $B/(2\lambda_L)$ . 
{\it This argument was not used in the derivation of Eq. (38)} and thus constitutes an independent derivation of Eq. (38).

Further corroborating evidence is provided by the fact that the magnetic field required to bring one of the components of the spin current to a stop has magnitude
$B_s$ given by\cite{sm}
\beq
B_s=-\frac{\hbar c}{4e\lambda_L^2}=\frac{\phi_0}{4\pi \lambda_L^2}
\eeq
which is essentially the expression for the lower critical field $H_{c1}$ of type II superconductors in the conventional theory of superconductivity\cite{tinkham}. This indicates that
superconductivity cannot exist unless spin current flows.

The most remarkable consequence of this physics is that the magnitude of the angular momentum of electrons in orbits of radius $2\lambda_L$ with velocity Eq. (38) is
\beq
l=m_e v_\sigma \times (2\lambda_L)=\frac{\hbar }{2}   .
\eeq
We will discuss this key point further in   later sections.
 
   \section{Relation between amount  of charge expelled and speed of the spin current }
   The magnitude of charge expelled and the maximum electric field $E_m$ are intimately related to the orbit expansion and spin current
   generation. The following simple argument
   shows this connection. The superfluid charge velocity in the Meissner effect is given by
   \beq
   v_s=-\frac{e}{m_ec}A=-\frac{e}{m_ec}\lambda_LB
   \eeq
and the magnitude of the charge current is given by
\beq
j=-n_s e v_s=\frac{n_s e^2}{m_e c}\lambda_LB=\frac{c}{4\pi\lambda_L}B
\eeq
where we have used Eq. (8). Using Eq. (30), Eq. (45) becomes
\beq
j=-\rho_- \frac{B}{E_m} c  .
\eeq
Eq. (46) can be interpreting as saying that the Meissner current is carried by the expelled charge density $\rho_-$ moving at speed
\beq
v_{\rho_-} = \frac{B}{E_m}  c
\eeq
It is natural to conclude that superconductivity will be destroyed when the speed $v_{\rho_-}$ reaches the speed of light, which requires (Eq. (42))
\beq
E_m=B_s=-\frac{\hbar c}{4e\lambda_L^2}
\eeq
and the excess charge density near the surface is, from Eqs. (8), (30), (38) and (48)
\beq
\rho_- = en_s\frac{\hbar}{4m_ec\lambda_L}=en_s\frac{v_\sigma^0}{c} .
\eeq

For the case where $H_{c1}=H_c$, that is at the crossover between type II and type I behavior, we can also express $\rho_-$ in terms of the
condensation energy per electron given by
\beq
\epsilon\equiv \frac{1}{n_s}  \frac{H_c^2}{8 \pi}
\eeq
as
\beq
\rho_-=en_s(\frac{2\epsilon}{m_e c^2})^{1/2}
\eeq
which shows that the charge expelled is a small fraction ($\sim 10^{-6})$ of the superfluid density.

The fact that $E_m$ is given by Eq. (48) also implies a simple relation between the electrostatic energy cost in setting up the charge separation and
the magnetic energy associated with the critical magnetic field $H_{c1}$. For a cylindrical geometry, the electrostatic energy cost 
per unit volume is
\beq
u_E=\frac{1}{2}\frac{E_m^2}{8\pi}
\eeq
which is half the maximum magnetic energy cost per unit volume
\beq
u_B=\frac{H_{c1}^2}{8\pi}
\eeq
in expelling the magnetic field, since $E_m=H_{c1}$. This is for the case of type II superconductors, where $H_{c1}\leq H_c$. Thus the
electrostatic energy cost is less than half the condensation energy density $H_c^2/8\pi$. It is also interesting to note that the
kinetic energy density of the spin current equals the electrostatic energy density near the surface where the electric field is $E_m$:
\beq
n_s \frac{1}{2}m_e v_\sigma^2=\frac{E_m^2}{8\pi}
\eeq
as can be seen from Eqs. (48), (38) and (8). We will shed some light into this relation in Sect. XI.

Finally, it should be pointed out that the relation Eq. (48) and resulting Eq. (49) follow directly from the requirement that the electrodynamic equations for the 
spin current be relativistically covariant. This somewhat lengthy derivation is given in Ref. \cite{electrospin}.

\section{Spin electrodynamics}

In a cylindrical geometry, the four-dimensional spin current is given by\cite{electrospin}
    \beq
J_\sigma(\vec {r},t)-J_{\sigma 0}=-\frac{c}{8\pi \lambda_L^2}(A_\sigma (\vec{r},t)-A_{\sigma 0} (\vec{r}))
\label{constitutive}
\eeq
with
\bmath
\beq
 J_\sigma(\vec{r},t)=(\vec{J}_\sigma(\vec{r},t),ic\rho_\sigma(\vec{r},t))
\eeq
\beq
 J_{\sigma 0} =(\vec{J}_{\sigma 0},ic\rho_{\sigma 0})
\eeq
\emath
and
\bmath
\beq
A_\sigma(\vec{r},t)=(\vec{A}_\sigma(\vec{r},t),i\phi_\sigma(\vec{r},t))
\eeq
\beq
A_{\sigma 0}(\vec{r})=(\vec{A}_{\sigma 0}(\vec{r}),i\phi_{\sigma 0}(\vec{r}))
\eeq
\emath
with
\bmath
\beq
\vec{A}_\sigma(\vec{r},t) =\lambda_L \vec{\sigma}\times \vec{E}(\vec{r},t)+\vec{A}(\vec{r},t)
\eeq
\beq
\vec{A}_{\sigma 0}(\vec{r}) =\lambda_L \vec{\sigma}\times \vec{E}_0(\vec{r})
\eeq
\emath
and
\bmath
\beq
  \phi_\sigma(\vec{r},t)=-\lambda_L\vec{\sigma}\cdot\vec{B}(\vec{r},t)+\phi(\vec{r},t)
  \eeq
  \beq
  \phi_{\sigma 0}(\vec{r})=\phi_0(\vec{r})
  \eeq
  \emath
$\vec{A}$ and $\phi$ are the magnetic vector potential and electric potential. $\vec{E}_0$ and $\phi_0$ are the electrostatic field and
  potential for a uniform charge density $\rho_0$ throughout the material, related to $\rho_-$ (Eq. (49)) by Eq. (29). 
     The current 4-vectors are given in terms of the velocity of the superfluid charge density per spin $en_s/2$, the velocity for each
  spin component $v_\sigma$, and the (excess) charge density $\rho_\sigma$  as
  \bmath
  \beq
  J_\sigma (\vec{r},t)=(\frac{en_s}{2}\vec{v}_\sigma(\vec{r},t),ic\rho_\sigma(\vec{r},t))
  \eeq
    \beq
  J_{\sigma 0}=(\frac{en_s}{2}\vec{v}_{\sigma 0},ic\rho_{\sigma 0})
  \eeq
    \emath
  with $\vec{v}_{\sigma 0}$ given by 
 \beq
    \vec{v}_{\sigma 0}=\vec{v}_\sigma(r<<R)=-\frac{c}{en_s}\rho_0 \vec{\sigma}\times\hat{r}
    \eeq
  
    The differential equations determining the behavior of all quantities are
    \bmath
\beq
\Box ^2 ( A_\sigma-A_{\sigma 0})=\frac{1}{\lambda_L^2}(A_\sigma -A_{\sigma 0})
\eeq
\beq
\Box^2 ( J_\sigma-J_{\sigma 0})=\frac{1}{\lambda_L^2}  (J_\sigma-J_{\sigma 0}) .
\eeq
\emath
with 
\beq
\Box^2=\nabla^2-\frac{1}{c^2}\frac{\partial^2}{\partial t^2}
\eeq
and $J_\sigma$ is given in terms of  $A_\sigma$ by Eq. (55). The equations for the charge sector only are simply obtained by defining the charge four-current and charge four-potential
\bmath
\beq
J_c=J_{\sigma=+1}+J_{\sigma=-1}
\eeq
\beq
A_c=(A_{\sigma=+1}+A_{\sigma=-1})/2
\eeq
\emath
and similarly for $J_{c0}$ and $A_{c0}$, and coincide with the equations given in Sec. V.
  
  The derivation of all these relations is given in Ref. \cite{electrospin}.
  
 \section{Rashba physics  and kinetic energy lowering}
 The interaction of the electron magnetic moment with an electric field $\vec{E}$  (spin-orbit interaction) is given by\cite{dirac}  
 \beq
 H_{s.o.}=-\frac{e\hbar}{4m_e^2 c^2} \vec{\sigma}\cdot(\vec{E}\times\vec{p})
 \eeq
 and a single electron Hamiltonian giving rise to this term to linear order is\cite{ac1,ac2,ac3,ac4}
  \bmath
     \beq
 H= \frac{1}{2m_e}(\vec{p}-\frac{e}{c}\vec{A}_\sigma)^2
 \eeq
  \beq
 \vec{A}_\sigma= \frac{\hbar}{4m_ec} \vec{\sigma} \times \vec{E}
 \eeq
 \emath
where we have used $\vec{\nabla}\times\vec{E}=0$. Unlike in other contexts\cite{ac1,ac2,ac3,ac4}, the quadratic
term arising from Eq. (66a) has real physical significance, as discussed in the next section.

The Spin Meissner effect can be understood as follows\cite{sm}: just as for the ordinary Meissner effect, we assume that the wavefunction in the
superconductor is rigid and hence $\vec{p}=0$ independent of the value of $\vec{A}_\sigma$. Using that $m_e\vec{v}_\sigma=\vec{p}-(e/c)\vec{A}_\sigma$
we obtain for the velocity of the electron of spin $\vec{\sigma}$
\beq
\vec{v}_\sigma=-\frac{e\hbar}{4m_e^2c^2}\vec{\sigma}\times\vec{E}
\eeq
Now electric fields in superconductors are screened over distances of order $\lambda_L$ according to the electrodynamic equations discussed in Sect. V.
In addition we have argued that the Meissner effect requires that carriers occupy orbits of radius $2\lambda_L$. The superfluid of density
$n_s$ $holes$ per unit volume carries a charge density $-en_s$, with $e$ the (negative) electron charge. Hence it moves in a compensating charge background
of density 
\beq
\rho=+en_s
\eeq
The electric field at the surface of a cylinder of radius $\vec{r}=2\lambda_L\hat{n}$ and charge density $\rho=en_s$ is
\beq
\vec{E}=2\pi e n_s\vec{r} = \frac{m_e c^2}{2e\lambda_L^2} \vec{r}=\frac{m_e c^2}{e\lambda_L} \hat{n}
\eeq
(where we have used Eq. (8)), with $\hat{n}$ the normal to the cylinder surface pointing outward. Replacing $\vec{E}$ in Eq. (67) yields
 \beq
   \vec{v}_\sigma^0=-\frac{\hbar}{4m_e \lambda_L} \vec{\sigma}\times \hat{n}
  \eeq
  in agreement with Eq. (38). Note that the correct sign of the spin current velocity (consistent with the force balance shown in Fig. 6) is obtained
  if the superfluid carriers are holes (Eq. (68)) rather than electrons.
  
  The Hamiltonian Eq. (66) to linear order yields
  \beq
  H=\frac{p^2}{2m_e} -\frac{\hbar}{4m_e\lambda_L} \vec{p}\cdot(\vec{\sigma}\times\hat{n})
  \eeq
using Eq. (69) for the electric field. Taking $\vec{p}=\hbar\vec{k}$ eq. (71) yields single-particle energy bands given by\cite{holecore}
\beq
\epsilon_{k\sigma}=\frac{\hbar^2}{2m_e}(k-\sigma \frac{q_0}{2})^2-\frac{\hbar^2 q_0^2}{8m_e}
\eeq
with $q_0=1/2\lambda_L$. This gives rise to two Rashba bands, 
with spin orientation perpendicular to both the momentum vector and the electric field (which is normal to the  
surface) and overall kinetic energy lowering of
$\hbar^2q_0^2/(4m_e)$ per charge carrier, as discussed in Ref. \cite{holecore}. The speed of the carriers resulting from Eq. (72)
\beq
v_{k\sigma}=\frac{1}{\hbar}\frac{\partial \epsilon_{k\sigma}}{\partial k}=\frac{\hbar}{m_e}(k-\sigma\frac{q_0}{2})
\eeq
correctly yields Eq. (38) for the spin current velocity.

\section{the phase of the superconducting electron, kinetic energy lowering and charge expulsion}

The orbital angular momentum of electrons in superconductors was found to be $\hbar/2$ (Eq. (43)). This result was found dynamically\cite{sm}, however its
form makes it clear that it has a topological origin. It is the minimum angular momentum corresponding to a $double$ $valued$ wave function, i.e.
to a dependence of the wavefunction on the azimuthal angle  $\phi$
\beq
\Psi(r,\phi)=f(r)e^{i\phi/2}
\eeq
which implies that the $single$ $electron$ wave function is   double valued. In other words, the phase $\theta $ of a superconducting electron 
  changes
by $\pi$ in going around a loop:
\bmath
\beq
\Psi(\vec{r})=|\Psi(\vec{r})| e^{i\theta(\vec{r})}
\eeq
\beq
\theta(\phi+2\pi)=\theta(\phi)+\pi   .
\eeq
\emath
Consequently, we argue that the fact that the flux quantization in superconductors is in units $hc/2e$ rather than $hc/e$ should  be understood 
 as arising from the phase condition
 Eq. (75b), or equivalently from the fact that the orbital angular momentum of the single electron is $\hbar/2$, 
  instead of from the fact that the charge of the Cooper pair is $2e$ as it is traditionally done.

Because electrons are paired in superconductors however the total
wavefunction is single-valued. The phase of  the  superconducting electron can be interpreted as the angular position on its $2\lambda_L$ orbit,
and it is the same for all electrons of spin $\sigma$ because of macroscopic phase coherence,
and opposite to the phase of the superconducting electrons of spin $-\sigma$.
Note that phase coherence between electrons is required to avoid collisions because the different $2\lambda_L$ orbits are highly overlapping.
This is shown schematically in Fig. 7.

 \begin{figure}
 \resizebox{8.5cm}{!}{\includegraphics[width=7cm]{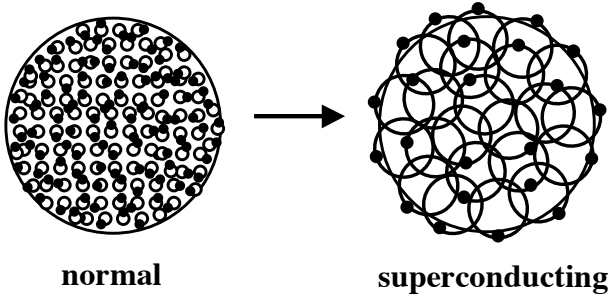}}
 \caption {Electronic orbits in the normal state have radius $k_F^{-1}$, of order of the ionic lattice spacing, and electronic orbits don't overlap.
 In the transition to superconductivity the orbits expand to radius $2\lambda_L$, several hundreds Angstrom, and they become highly overlapping.
 The black dots on the orbits denote the instantaneous position of the electron i.e.  its   `phase', which is random in the normal state and coherent in the superconducting state.
 Electrons of opposite spin rotate in opposite directions.
   }
 \label{figure2}
 \end{figure}

It is natural to conclude  that this double-valuedness found here at the mesoscopic level ($2\lambda_L$ scale) and associated angular momentum
$\hbar/2$ reflects an intrinsic property of the electron wave function at all length scales\cite{double}. The kinetic energy of a classical particle with
angular momentum $L$ is 
\beq
E_{kin}=\frac{L^2}{2mr^2}   .
\eeq
If the angular momentum $L$ has a fixed non-zero value, Eq. (76) implies that the kinetic energy is lowered when the wavefunction
expands (increasing $r$). This would provide a general physical explanation for `quantum pressure', i.e. the tendency of quantum particles to
expand their wavefunction to lower their kinetic energy. We return to this point in the next section.

In the present context, the charge expulsion,  kinetic energy lowering, orbit expansion and their relation with the phase condition Eq. (75b) can be understood as follows. The total energy
of an electron is the sum of kinetic energy Eq. (76) and  potential energy:
\beq
E_{pot}= \frac{e^2 E_m^2}{8m_ec^2} r^2   .
\eeq
This term is   the square of the spin-orbit vector potential from the Hamiltonian Eq. (66),
 using the electric field given by Eq. (69) and the expression for $E_m$ Eq. (48). 
 The sum of kinetic and potential energies, assuming $L=\hbar/2$ in Eq. (76) as determined by
 the phase condition Eq. (75b) is
 \beq
 E_{tot}=E_{kin}+E_{pot}=\frac{\hbar^2}{8m_er^2}+\frac{e^2 E_m^2}{8m_ec^2} r^2
 \eeq
 and minimization with respect to $r$ ($\partial E_{tot}/\partial{r}=0$)  yields for the radius of lowest energy
 \beq
 r=2\lambda_L  
 \eeq
 as expected.
 
 The physical origin of the potential energy Eq. (77) is the following: as the orbits expand the charge density buildup associated with the
 accompanying charge expulsion is $\rho_-$, Eq. (49), which generates an electric field at distance $r$ (in cylindrical geometry)
 \beq
 \vec{E}_s(\vec{r})=2\pi|\rho_-|\vec{r}=\frac{E_m}{2\lambda_L}\vec{r}
 \eeq
 using Eq. (30). Note that this ``screened'' electric field is much smaller than the ``bare'' electric field Eq. (69) originating in the full ionic
 charge density $|e|n_s$. The proportionality factor is $(v_\sigma^0/c)$ (Eq. (49)), of order $10^{-6}$. This small electric field gives rise to
 an electrostatic energy cost per unit volume
 \beq
 U_{elec}=\frac{E_s(r)^2}{8\pi}=\frac{E_m^2}{8\pi}\frac{r^2}{(2\lambda_L)^2}
 \eeq
 hence an electrostatic energy per superfluid carrier (using Eq. (8))
 \beq
 E_{pot}=\frac{1}{n_s}U_{elec}= \frac{e^2 E_m^2}{8m_ec^2} r^2   .
\eeq
 in agreement with Eq. (77).
The electrostatic energy for  the equilibrium radius $r=2\lambda_L$ is 
 \beq
 E_{pot}=\frac{1}{n_s}\frac{E_m^2}{8\pi}
 \eeq
 which properly represents the electrostatic energy cost per superfluid carrier originating in the charge expulsion that builds up the internal field $E_m$. Thus, the expansion of orbits from the microscopic
 radius $k_F^{-1}$ to the mesoscopic radius $2\lambda_L$ is understood as arising from  the competition of  lowering of kinetic energy Eq. (76) 
 with the electron angular momentum fixed at $L=\hbar/2$ and cost in electrostatic potential energy Eq. (77) originating in charge expulsion.   

\section{Discussion}

\begin{figure}
\resizebox{5.5cm}{!}{\includegraphics[width=7cm]{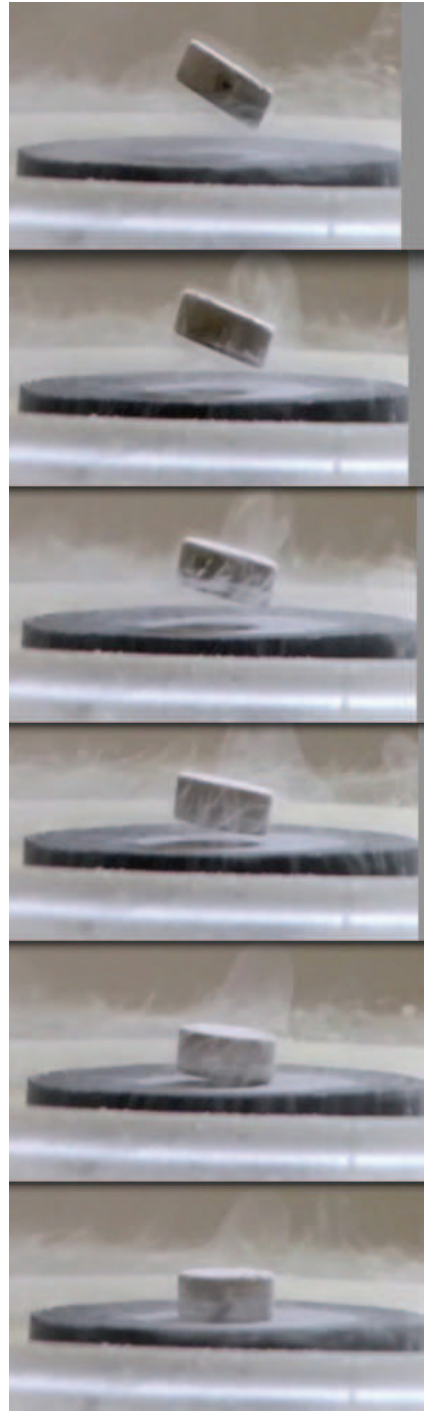}}
  \caption{Spontaneous lifting of a magnet resting on top of a metal  being cooled into the superconducting state (bottom to top).
  The force  that pushes the magnet up against gravity originates in the 
   force driving the electronic orbits in the superconductor to expand, namely the kinetic energy lowering with increasing orbit size
  of electrons constrained to have angular momentum $\hbar/2$ at all length scales.}
\end{figure}

Figure 8 displays the fundamental physics of superconductors, that we propose is a direct consequence of kinetic energy lowering. 
In this paper we have pointed out that there is a fundamental connection between kinetic energy lowering, the Meissner effect, charge expulsion,
rotational zero-point motion,  and superconductivity.

If superconductivity is kinetic energy driven, it should incur a cost in potential energy. Within our theory, this cost is very apparent: it is the electrostatic energy cost
generated by charge expulsion (Fig. 2).   The charge expulsion is associated with the expansion of the electronic orbits. The increased diamagnetic susceptibility of 
superconductors compared to normal metals (Meissner effect)  indicates that the orbits of the charge carriers increase their radius as the metal goes superconducting,
or, in other words, that their wavefunction increases its spatial extent, as described by the Larmor diamagnetic susceptibility
\beq
\chi_{Larmor}=-\frac{e^2}{4m_e c^2}<r_\bot ^2>   
\eeq
with $r_\bot$ the radial coordinate in a plane. In quantum mechanics, expansion of the wave function, i.e. increasing $<r_\bot ^2>$, is associated with
lowering of the expectation value of the kinetic energy
\beq
E_{kin}=<-\frac{\hbar^2}{2m}\nabla^2>
\eeq
as can be seen from the fact that a lower bound for the kinetic energy of an electron is \cite{lieb}
 \beq
E_{kin}\geq \frac{3}{5}    (6\pi^2)^{2/3 }\frac{\hbar^2}{2m_e} \frac{\int d^3r \rho(\bold{r})^{5/3} }{ \int d^3r \rho(\bold{r})}
 \eeq
 with $\rho(\br)$ the electron  density. For a wavefunction occupying a radial extent $\bar{r}$, $\rho(\br)\sim 1/\bar{r}^3$, 
 $\rho(\br)^{5/3}\sim 1/\bar{r}^5$ and 
 the right side of Eq. (83) $\sim 1/\bar{r}^2$. Alternatively, the fact that the kinetic energy is lowered as the spatial extent of the wavefunction increases is seen 
 from Heisenberg's uncertainty principle.

In accordance with Bohr's correspondence principle, this connection can also be understood   classically. Eq. (84) follows from Faraday's law for a circular
orbit of radius $r$, with $r^2$ replacing $<r_\bot ^2>$. For a particle of mass $m$ and angular momentum $L$ the kinetic energy in such an orbit is
Eq. (76),  hence the kinetic energy is lowered as the orbit expands {\it for fixed L}. Taking $L=\hbar$, that relation becomes
\beq
E_{kin}=\frac{\hbar^2}{2mr^2}
\eeq
which coincides with the quantum-mechanical value for the atomic kinetic energy with the wavefunction Eq. (2), with $\bar{r}$ replacing $r$. This classical argument can be used to understand the relation between kinetic energy and orbit radius in the Bohr atom, for which all orbits have nonzero
angular momentum. In Schr\"odinger theory however, `quantum pressure', i.e. the tendency of electrons to expand their wavefunction to lower their kinetic
energy, exists also for states of zero angular momentum, as expressed e.g. by Eq. (85) or by the uncertainty principle. This physics does not have a classical counterpart.

It is thus remarkable that for superconductors we have found that kinetic energy lowering and orbital expansion are intimately tied to the fact
that the orbital angular momentum of the electron is constrained to be $\hbar/2$. This suggests that our classical interpretation is   more
correct than the quantum-mechanical one. Namely, that electronic `quantum pressure' originates always in finite angular momentum rather than the
uncertainty principle, and since quantum pressure is ubiquitous, that states of zero angular momentum for the electron don't exist. 
{\it The angular momentum of the electron is bounded from below by
$\hbar/2$ due to the topological constraint eq. (75b).}

Another way to put it: in conventional (Schr\"odinger) quantum mechanics, there is no ``rotational zero point motion'': the fact that the azimuthal angle is
constrained to the finite angular interval $0^\circ$  to $360^\circ$ does not raise the kinetic energy of a particle of angular momentum zero. But we
have found that superconductors have rotational zero point motion at the macroscopic level, and superconductors are macroscopic quantum
objects. This suggests that very generally there should be rotational zero point motion at the microscopic level also. This is 
incompatible with states of zero orbital angular momentum.

Electrons have intrinsic angular momentum $\hbar /2$ (spin). This can be understood as originating in circular orbits of radius
\beq
r_q=\frac{\hbar}{2m_e c}
\eeq
with electrons moving at the speed of light\cite{hestenes}. If the orbit expands to radius $2\lambda_L$, keeping the angular momentum fixed,
the speed is reduced by a factor $v_\sigma^0/c=r_q/(2\lambda_L)$ yielding the spin current speed Eq. (38). 
Note  that this factor gives also the ratio between the expelled charge density $\rho_-$ and the `bare' charge density $en_s$. 
All this suggests that the orbits in the superconductor are
a mesoscopic image at length scale $2\lambda_L$ of the spinning electron motion at the scale Eq. (88). Furthermore, it suggests that the electron has minimum angular momentum
$\hbar/2$ at all length scales. As discussed in the previous section, this would be the case if the orbital wavefunction for the electron is double-valued. 
The possibility that the wavefunction of the electron is double-valued has been considered in the past 
by Eddington, Schr\"odinger and Pauli \cite{d1,d2,d3}.

David Hestenes has proposed long ago that the electron spin should be interpreted as an $orbital$ angular momentum\cite{hestenes}. 
He has pointed out that the Schr\"odinger equation should be regarded as describing an electron in an eigenstate of spin, and that spin
should be interpreted as a dynamical property of electron motion rather than an internal angular momentum. Furthermore, he has pointed out that
Heisenberg's uncertainty relations follow naturally from constraining the electron to have non-zero orbital angular momentum. These concepts
appear to be closely related to the physics discussed here.

In addition, the fermion anticommutation relations can be interpreted as arising naturally from the double-valuedness of the electron wave function
and the resulting phase condition Eq. (75b).  This is because the process of interchanging two fermions is topologically equivalent to one electron going around
the other in a loop, thus picking up a $(-)$ sign according to Eq. (75b). This interpretation of the origin of fermion anticommutation relations is
discussed by Feynman\cite{feynman,f2}. If so, the stability of matter, which according to Ref. \cite{lieb} results from the combined effect of
`quantum pressure' (e.g. Eq. (86)) at the single electron level and `Pauli pressure' originating from fermion anticommutation relations can instead be
uniquely ascribed to the double-valuedness of the electron wave function Eq. (75b).

In summary, the physics of superconductors discussed in this paper leads us to conclude that angular momentum plays an even more central
role in quantum mechanics than conventionally assumed. Namely, that the fundamental origin of quantum pressure, i.e. the 
kinetic energy of quantum confinement, is $always$ non-zero angular momentum, and that 
 the quantum $phase$ of a particle can be interpreted as arising from a rotational degree of freedom. This is, after all, not very
surprising given that it is $\hbar \neq 0$ that gives rise to quantum mechanics, and $\hbar$ has units of angular momentum.
 
 Concerning superconductivity, it remains to show how electrons in overlapping $2\lambda_L$ orbits maintain their phase coherence, for which it will be
  necessary to take into account scattering processes between pairs in a BCS-like wavefunction. Concerning other
problems, we have proposed that this physics  gives rise to a spin current 
ground state for aromatic ring molecules[69],  and there are hints that the concepts discussed here could have  implications for the    interpretation of the Dirac equation and  the quantum Hall effect.

In his talk on a meeting commemorating the 50-th anniversary of BCS theory, Steven Weinberg said\cite{reductionist} 
``I think that the single most important thing accomplished by the theory of John Bardeen, Leon Cooper, and Robert Schrieffer (BCS) was to show that superconductivity is not part of the reductionist frontier''. However, the results presented in this paper suggest that the correct understanding of superconductivity could
have a profound effect on the reductionist frontier, by requiring a reinterpretation of the origin of quantum pressure and a 
reformulation of conventional quantum mechanics to describe
an intrinsic  double-valuedness of the electron wave function\cite{double,double2}. The fact that there are serious problems with the conventional understanding  of
quantum mechanics has been pointed out emphatically by A.V. Nikulov\cite{nikulov}.

 \acknowledgements The author is grateful to Congjun Wu for stimulating discussions.

\end{document}